\documentclass[pre,twocolumn,superscriptaddress]{revtex4-1}
\usepackage{graphicx}
\usepackage{amsmath,amssymb}

\newcommand{\Tr}{{\rm Tr}}

\newcommand{\be}{\begin{equation}}
\newcommand{\ee}{\end{equation}}
\newcommand{\tn}{\widetilde{N}}

\begin{document}

\title{Quantum transport in chaotic cavities with tunnel barriers}
\author{Lucas H. Oliveira}
\affiliation{Instituto de F\'isica, Universidade Federal de Uberl\^andia, 38408-100, Brazil}
\author{Pedro H. S. Bento}
\affiliation{Instituto de F\'isica, Universidade Federal de Goiás, 74690-900, Brazil}
\author{Marcel Novaes}
\affiliation{Instituto de F\'isica, Universidade Federal de Uberl\^andia, 38408-100, Brazil}

\date{\today}

\begin{abstract}

We bring together the semiclassical approximation, matrix integrals and the theory of symmetric polynomials in order to solve a long standing problem in the field of quantum chaos: to compute transport moments when tunnel barriers are present and the number of open channels, $M$, is small. In contrast to previous approaches, ours is non-perturbative in $M$; instead, we arrive at an explicit expression in the form of a power series in the barrier's reflectivity, whose coefficients are rational functions of $M$. For general moments we must require that the barriers are equal and time reversal symmetry is broken, but for conductance we treat the general situation. Our method accounts for exponentially small non-perturbative terms that were not accessible to previous semiclassical approaches. We also show how to include more than two leads in the system.
\end{abstract}

\maketitle

\section{Introduction}

We consider quantum transport through ballistic systems with chaotic dynamics, like a two dimensional electron gas confined in mesoscopic cavity which is attached to leads \cite{nazarov}. This has long been a prime testing ground for ideas from quantum chaos, the study of the interplay between unpredictability due to dynamics and quantum uncertainty \cite{qc1,qc2,qc3,qc4}. One of the main findings was universality: average observables are insenstive to system's details and, ignoring spin, depend only on whether time reversal symmetry (TRS) is present or not \cite{haake}.

We consider initially two leads, supporting $N_1$ and $N_2$ open channels. Quantum scattering in this context is described by the $M\times M$ unitary $S$ matrix relating incoming to outgoing quantum amplitudes, where $M=N_1+N_2$. Let $t$ be the block from $S$ which contains the transmission amplitudes between the leads. Then the hermitian matrix $t^\dagger t$ encodes the relevant transport properties that characterize the electrical current as a funcion of time and energy \cite{landauer,buttiker}. The (dimensionless) conductance of the system is given by $g=\Tr(t^\dagger t)$. It is a widely fluctuating function of the energy and one considers its average value $\langle g\rangle$, within a range of energies which is classically small but large in the quantum scale. Conductance fluctuations involve $\langle[\Tr(t^\dagger t)]^2\rangle$ and shot-noise \cite{shot} involves $\langle \Tr[(t^\dagger t)^2]\rangle$.

When the leads connecting the cavity to the outside world are ideal, i.e. perfectly transmitting, it is well known that the average conductance is given by $N_1N_2/M$ when time-reversal symmetry is absent and by $N_1N_2/(M+1)$ when it is present (the difference between these two results is known as the weak localization correction, or enhanced back-scattering). A more realistic setting is to assume the presence of tunnel barriers in the leads \cite{bar1,bar2,bar3,bar4}, so that channel $i$ has an associated tunnelling rate $\Gamma_i$, with $\Gamma_i = 1$ being the ideal case. We shall assume for simplicity that all such rates are equal within each lead, but the leads may be different, so $\Gamma_j$ characterizes lead $j$. Even with this simplification, calculations are much more challenging than in the ideal case (we review previous results in the next Section). 

In this work we avail ourselves of recent progress in the semiclassical approach to this problem, in which matrix elements of $S$ are written as sums over classical trajectories of complex numbers whose phase is proportional to the action. We build a matrix integral which encodes the semiclassical diagrammatic theory when tunnel barriers are present in all leads (which may be in arbitrary number). In this way we obtain, for both universality classes, expressions to the average conductance which are power series in the reflectivities $(1-\Gamma_1)$ and $(1-\Gamma_2)$ with coefficients that are rational functions of $M$. Our results are thus valid in the extreme quantum limit, in contrast to previous approaches which were perturbative in the parameter $1/M$. When time reversal symmetry is absent and $\Gamma_2=\Gamma_1$, we are also able to treat higher transport moments, i.e. general symmetric polynomials in the eigenvalues of $t^\dagger t$.  

In Section II, we review previous results and present some of our own. In Section III we develop our semiclassical matrix integral for computing the average conductance, both for  systems with and without TRS. The solution of the integral and presented in Section IV. Higher transport moments, like conductance variance and shot-noise, are treated in Section V. We conclude in Section VI. The Appendix reviews some combinatorial background.

\section{Some Results}

Brouwer and Beenakker considered the conductance problem in the presence of barriers within a random matrix theory (RMT) formulation, which forgoes dynamical details and treats the $S$ matrix statistically \cite{rmt}. They obtained the distribution of $g$ for a single channel \cite{BB2} and then, after introducing a diagrammatic method to perform averages over the unitary group, showed that for large $M$ its average value is
\be\label{bb} \langle g\rangle=\frac{\tn_1\tn_2}{\widetilde{M}}
+\left(1-\frac{2}{\beta}\right)\frac{\tn_1\tn_2\Gamma_1\Gamma_2(N_1+N_2)}{\widetilde{M}^3}+\cdots,\ee
where \be \tn_i=\Gamma_iN_i\ee is an effective number of channels in lead $i$ and
\be \widetilde{M}=\tn_1+\tn_2.\ee
The Dyson parameter $\beta$ in the above equation equals $1$ when TRS is present, and $2$ when it is broken. 

The variance of the conductance, $\langle ({\rm Tr}(t^\dagger t)-g)^2\rangle$, was also treated in \cite{BB}. Similar work was done in \cite{ramos,ramos2}, where the leading terms for the average shot-noise (quantum fluctuations in the electric current due to granularity of charge), proportional to $\langle g-{\rm Tr}(t^\dagger t)^2\rangle$, were obtained.

\begin{figure*}[t]
\includegraphics[scale = 0.2]{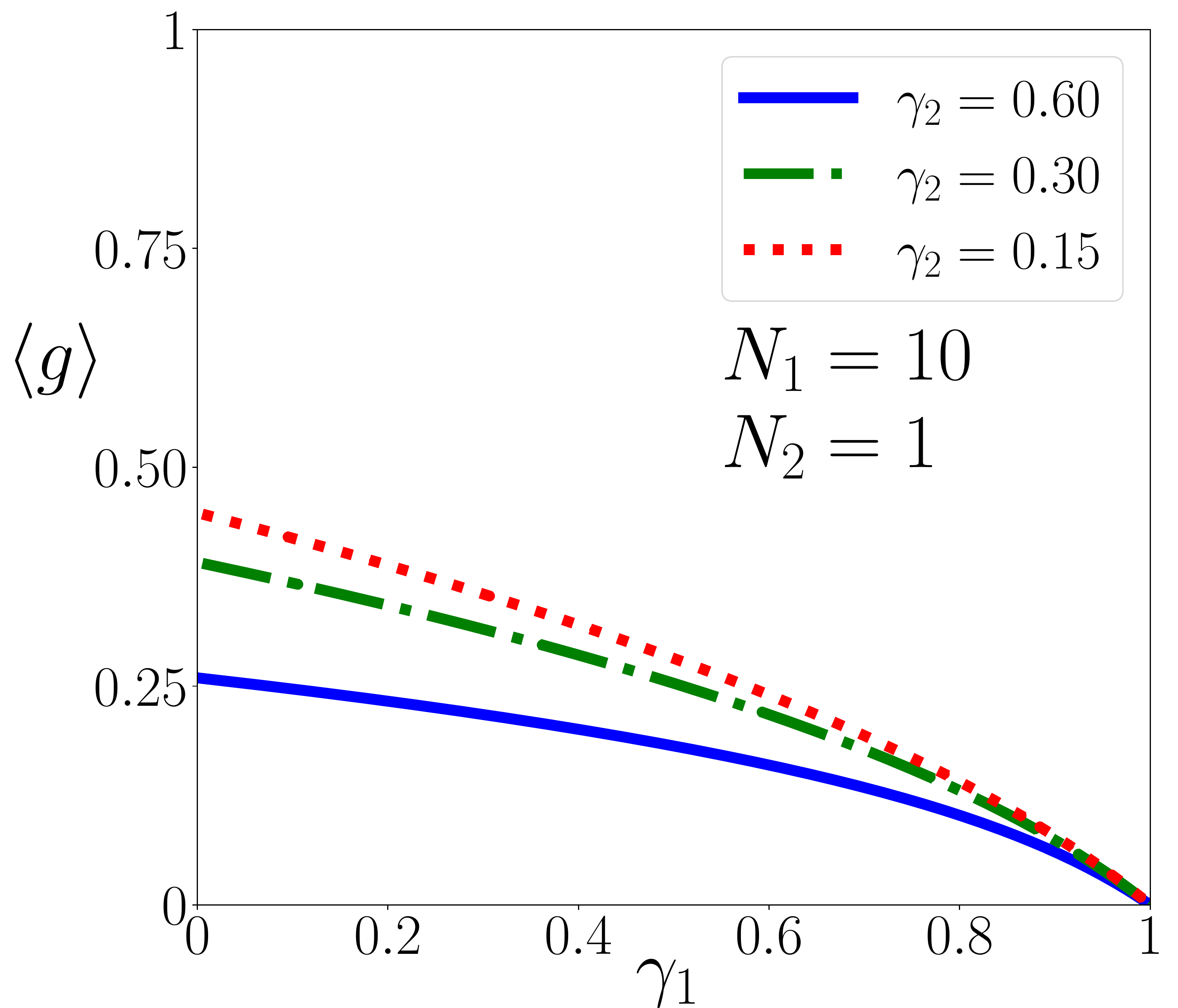}\hspace{0.1cm}\includegraphics[scale = 0.2]{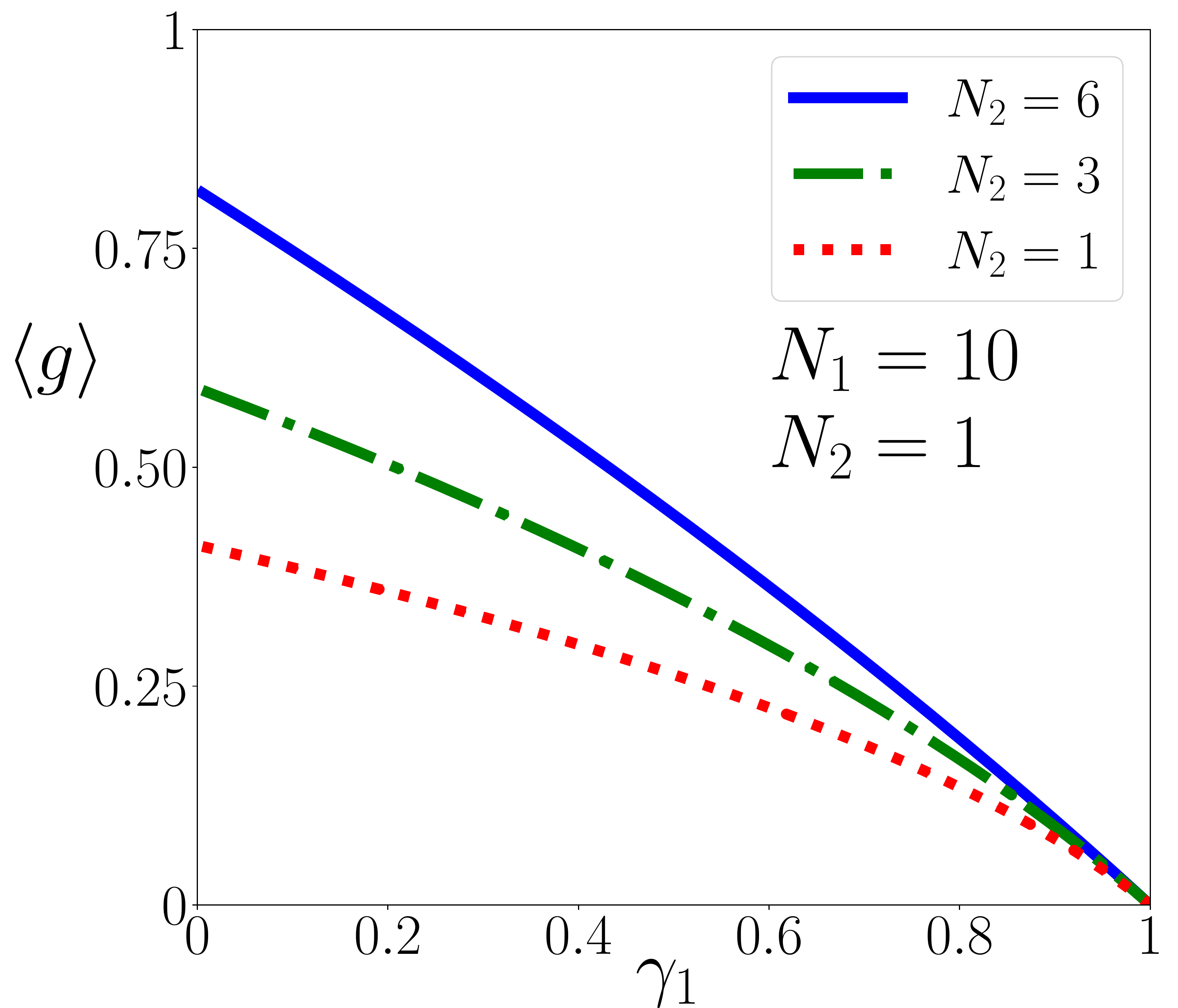}\\
\includegraphics[scale = 0.2]{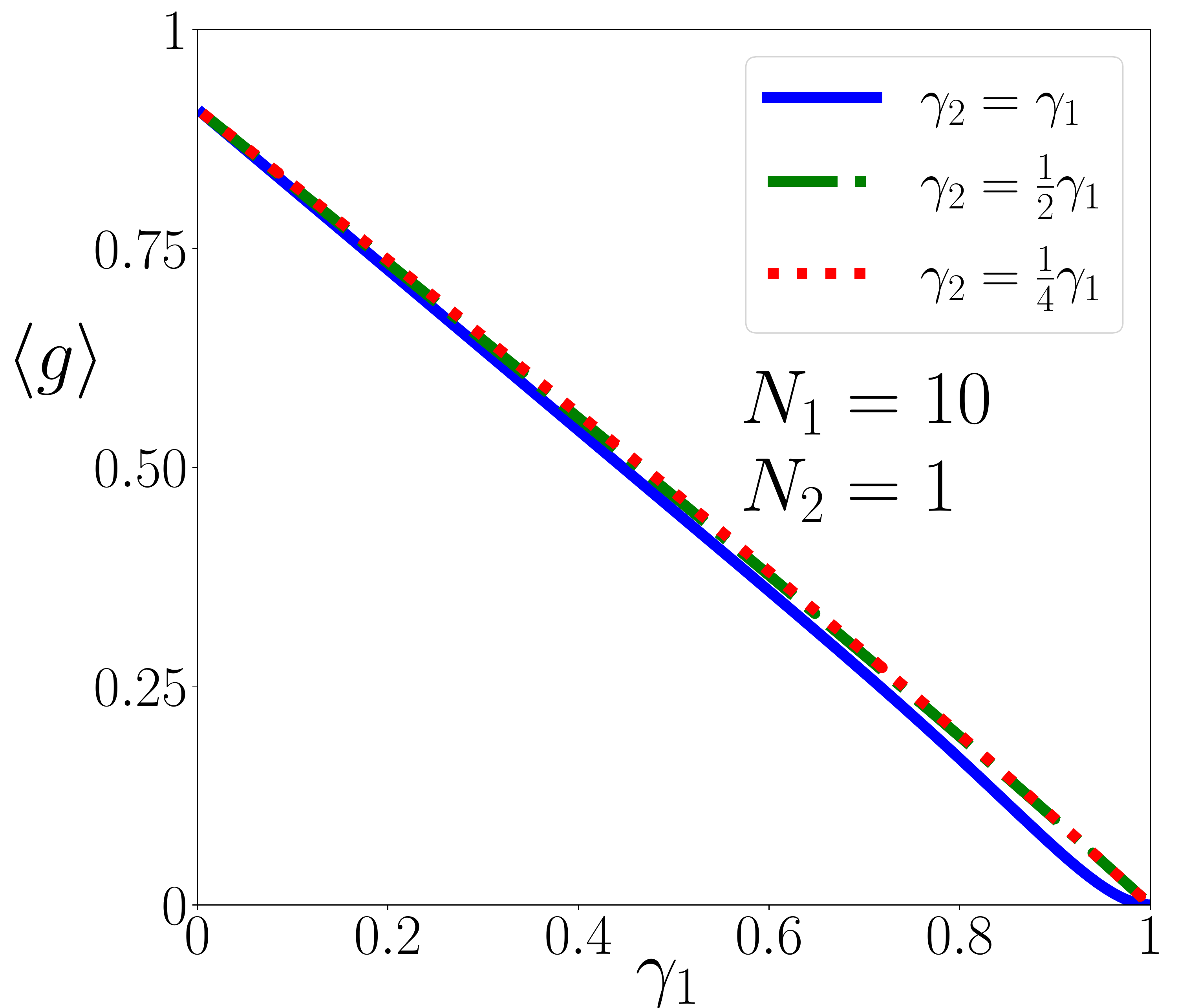}\hspace{0.1cm}\includegraphics[scale = 0.2]{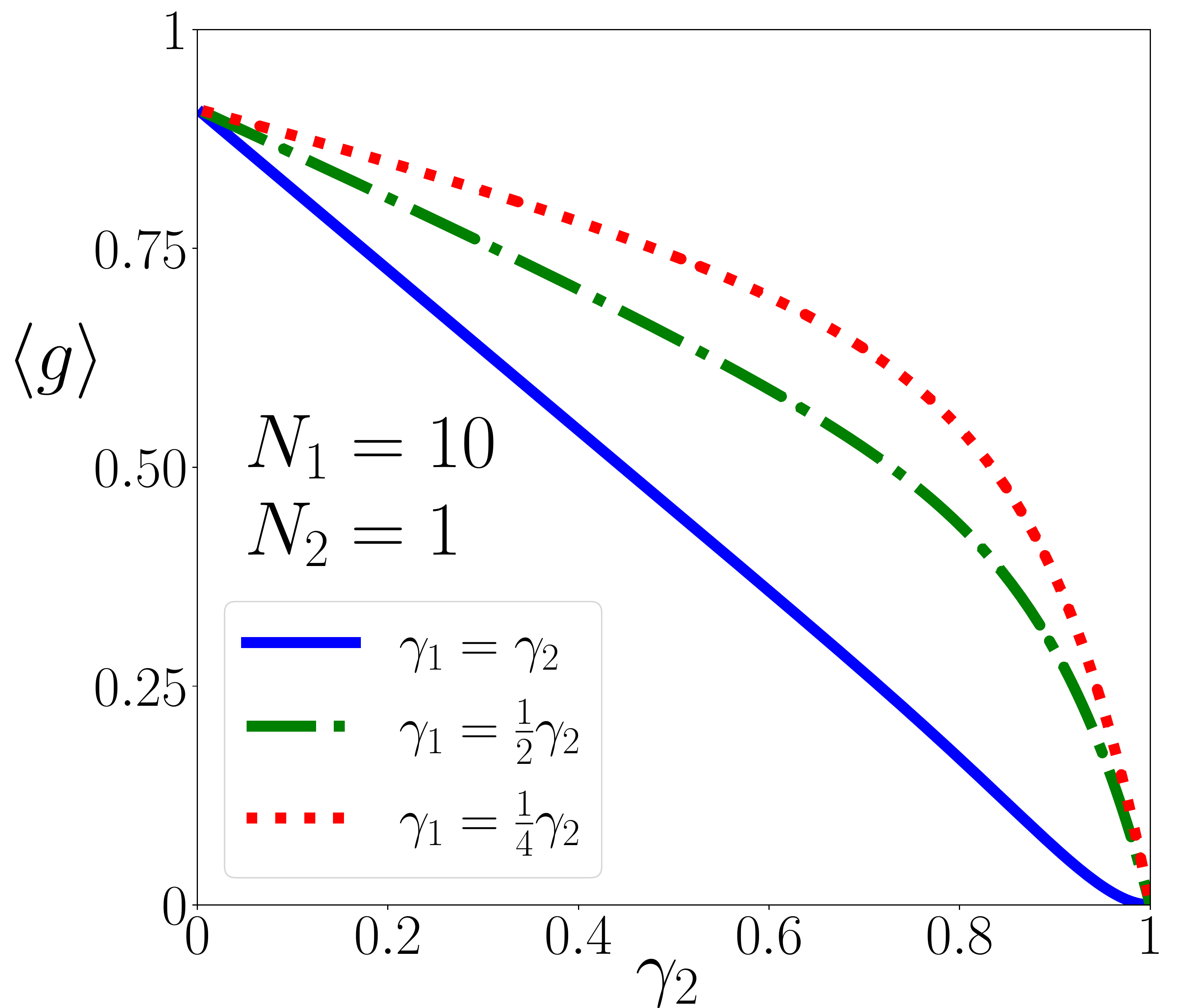}
\caption{(color online) Average conductance for broken time reversal symmetry, as a function of the reflectivities, in several different regimes.}
\end{figure*}

The above results are perturbative in $M$ and exact in $\Gamma_i$. Work by Kanzieper at. al. \cite{kanz1,vidal,kanz2} opened the possibility of results that are valid for finite $M$ by computing the RMT distribution of reflection eigenvalues, but only if the larger lead (lead 2, say) is kept ideal. Using that result and introducing the reflectivities
\be \gamma_i=1-\Gamma_i,\ee Rodr\'iguez-P\'erez and coworkers obtained \cite{perez} a formula for transport moments as a power series in $\gamma_1$ with coefficients that are rational functions of channel numbers. In the case of conductance, 
\be\label{gperez} \frac{\langle g\rangle_{\gamma_2=0}}{N_1N_2}=\frac{1}{M}-\frac{\gamma_1 N_2}{M^2-1}-\frac{\gamma_1^2N_2(MN_1-2)}{(M^2-1)(M^2-4)}+\cdots\ee The more general problem when both leads have tunnel barriers seems to be out of the reach of the methods used in \cite{kanz1,vidal,kanz2}.

The objective of the present work is to close that gap and find an expression for the average conductance in the presence of tunnel barriers in both leads, valid in the deep quantum regime when $M$ is small. We do not rely on RMT, however, but on the semiclassical approximation, which has a powerful diagrammatic formulation, developed in the ideal case by Haake and collaborators \cite{essen3,essen4,essen5}, building on previous work by Sieber and Richter \cite{sieber1,sieber2}. These two theories are expected to be in agreement for generic systems, and this has been fully established in the ideal case \cite{greg1,greg2,greg3,matrix,trs}. The diagrammatic rules were later adapted to account for tunnel barriers \cite{whitney,kuipers}, and perturbative calculations in $M$ were performed to recover the first few terms in (\ref{bb}) and (\ref{gperez}) for the conductance. Conductance variance and shot-noise were treated in \cite{jacquod}, while higher moments were investigated in \cite{kuipersrichter}. 

The semiclassical approach was recently successful \cite{pedro} in providing finite $M$ results but it was restricted to broken time reversal symmetry and, just like within RMT, lead 2 had to be kept ideal. The expressions obtained were in agreement with the ones from RMT, but considerably simpler. Conductance, for instance, was shown to be given by
\be \langle g\rangle_{\gamma_2=0}=\frac{\tn_1N_2}{M}\sum_{m=0}^\infty \frac{\gamma_1^m}{m+1}\sum_{k=0}^{m} \frac{(N_1+1)^{m-k}(N_1-1)_k}{(M+1)^{m-k}(M-1)_k},\ee
where $(M)^k=M(M+1)\cdots (M+k-1)$ and $(M)_k=M(M-1)\cdots (M-k+1)$ are the usual rising and falling factorials. 

In the present work we are able to include a tunnel barrier in lead 2 as well. We obtain explicit expressions for transport moments as power series in $\gamma_1$ and $\gamma_2$, with coefficients that are rational functions of channel numbers and depend on several concepts from combinatorics and representation theory. 

When time-reversal symmetry is broken, the leading terms for the average conductance are:
\begin{widetext}
\be \langle g\rangle=\frac{N_1N_2}{M}-\frac{N_1N_2(N_1\gamma_2+N_2\gamma_1)}{M^2-1}-\frac{N_1N_2(\gamma_1-\gamma_2)[N_1N_2M(\gamma_1-\gamma_2)+2(N_1\gamma_2-N_2\gamma_1)]}{(M^2-1)(M^2-4)}+\cdots\ee 

In Figure 1 we see the average conductance for systems with broken TRS in different regimes. As is to be expected, it is a decreasing function of $\gamma_1,\gamma_2$ and an increasing function of $N_1,N_2$. When one lead is much larger than the other, $\langle g\rangle$ is almost insensitive to the barrier in the smaller lead, but has a nontrivial dependece on the barrier in the larger one.

Rather curiously, when $\gamma_1=\gamma_2$ it seems that all orders higher than the first vanish identically (as far as we can compute), except for the $M$th order, and we conjecture that
\be\label{gu} \langle g\rangle_{\gamma_2=\gamma_1}=\frac{N_1N_2}{M}-\frac{N_1N_2M\gamma_1}{M^2-1}+\frac{N_1N_2\gamma_1^M}{M(M^2-1)}.\ee
The last term in the above expression is exponentially small in the large-$M$ regime and is missing from all previous semiclassical approaches (if this term is omitted, the resulting expression for $\langle g\rangle$ becomes negative in some regimes).

When time reversal symmetry is present, the leading terms are 
\begin{align} \langle g \rangle=&\frac{N_1N_2}{M+1}-\frac{N_1N_2((N_1+1)\gamma_2+(N_2+1)\gamma_1)}{M(M+3)}\nonumber\\&-N_1N_2\frac{[N_1N_2(M+1)+M^2-3](\gamma_1-\gamma_2)^2+2(\gamma_1^2-\gamma_2^2)(N_1-N_2)-4\gamma_1\gamma_2}{(M-1)M(M+3)(M+5)}+\cdots\end{align}
In particular, when $\gamma_1=\gamma_2$ we get
\be \langle g\rangle_{\gamma_2=\gamma_1}=\frac{N_1N_2}{M+1}-\frac{N_1N_2(M+2)\gamma_1}{M(M+3)}+\frac{4N_1N_2\gamma_1^2}{(M-1)M(M+3)(M+5)}+\cdots\ee
\end{widetext}

In Figure 2 we see the average conductance for systems with intact TRS, in the same regimes as Figure 1. The differences between the two universality classes are not very pronounced.

\begin{figure*}[t]
\includegraphics[scale = 0.2]{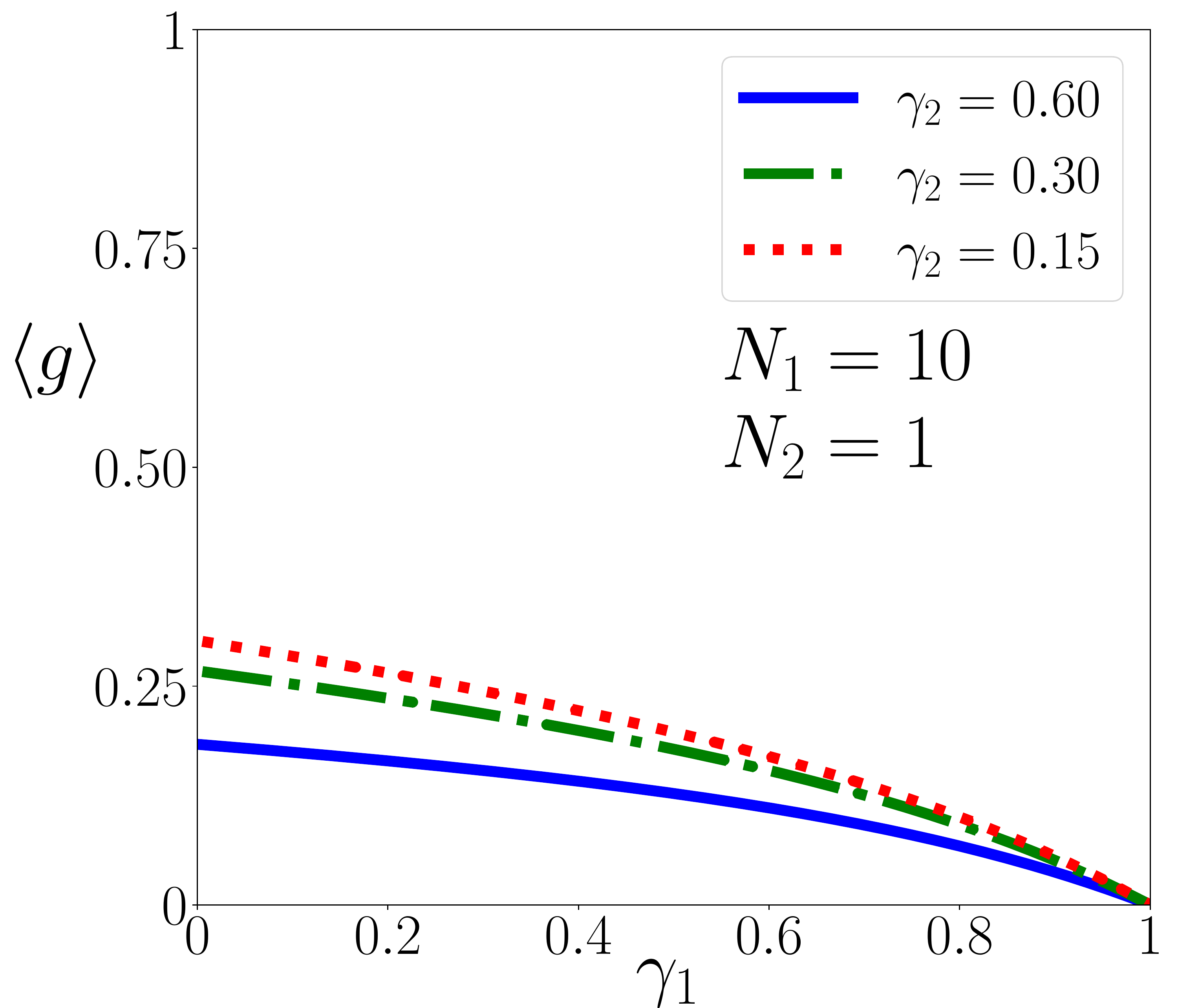}\hspace{0.1cm}\includegraphics[scale = 0.2]{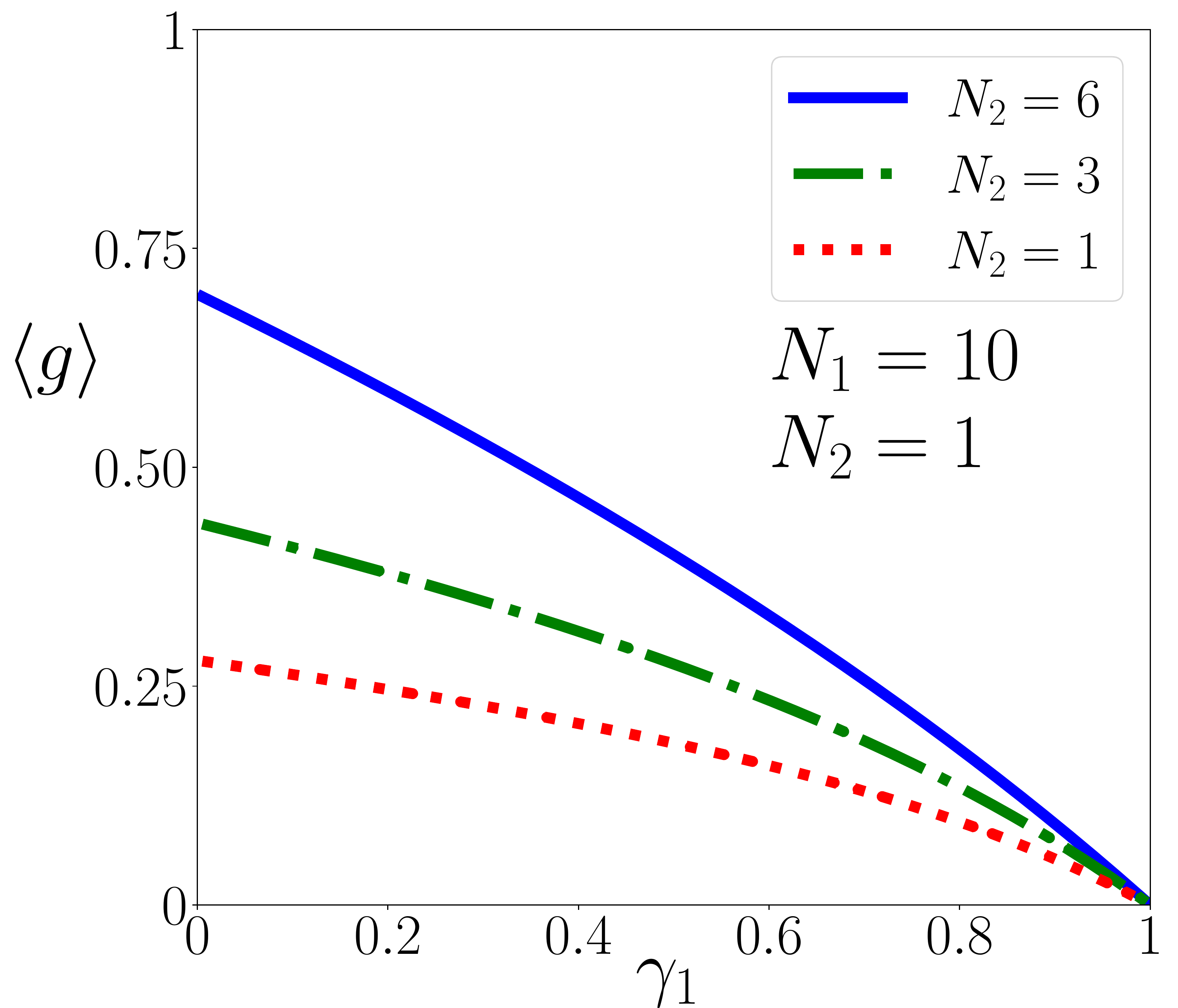}\\
\includegraphics[scale = 0.2]{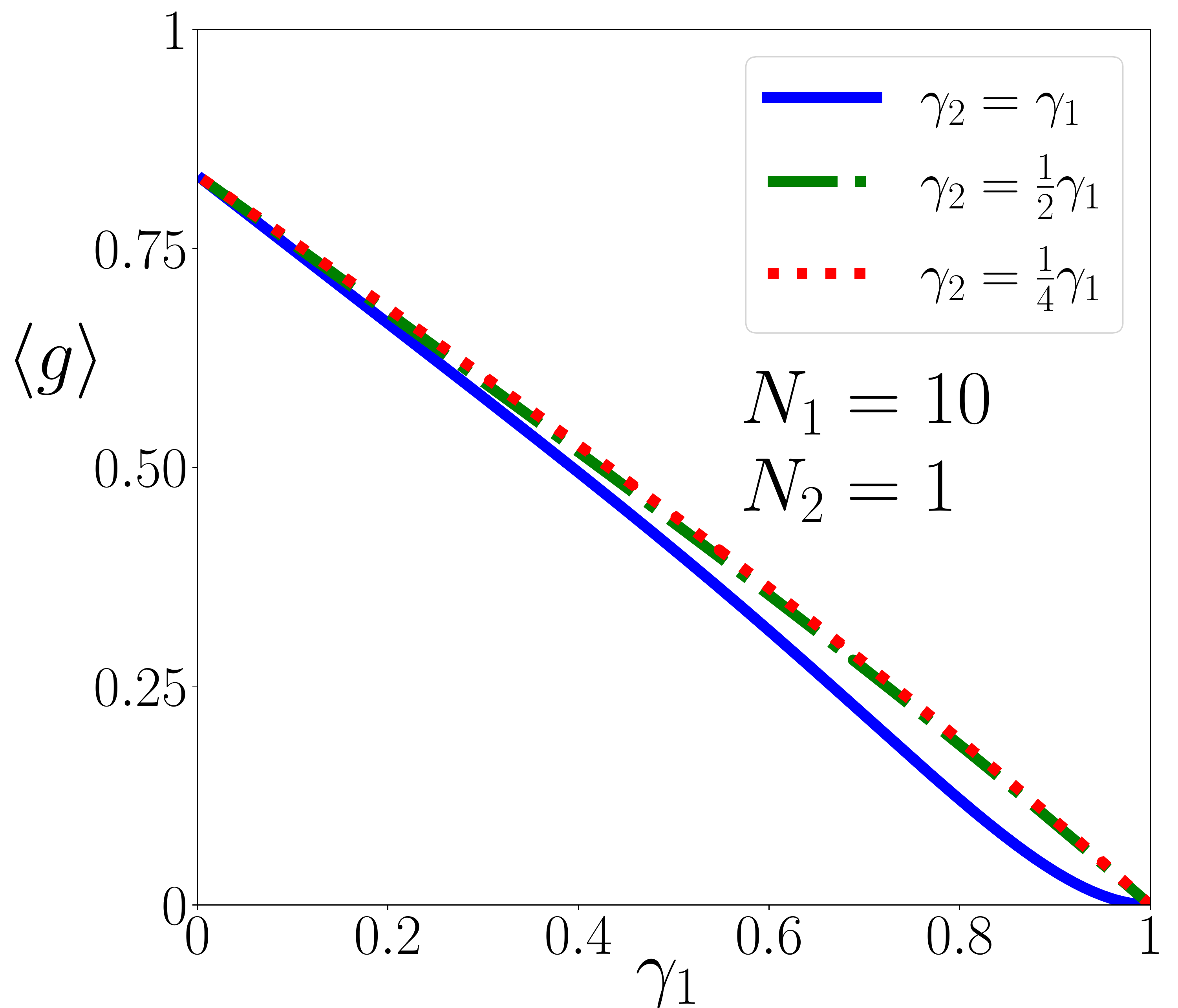}\hspace{0.1cm}\includegraphics[scale = 0.2]{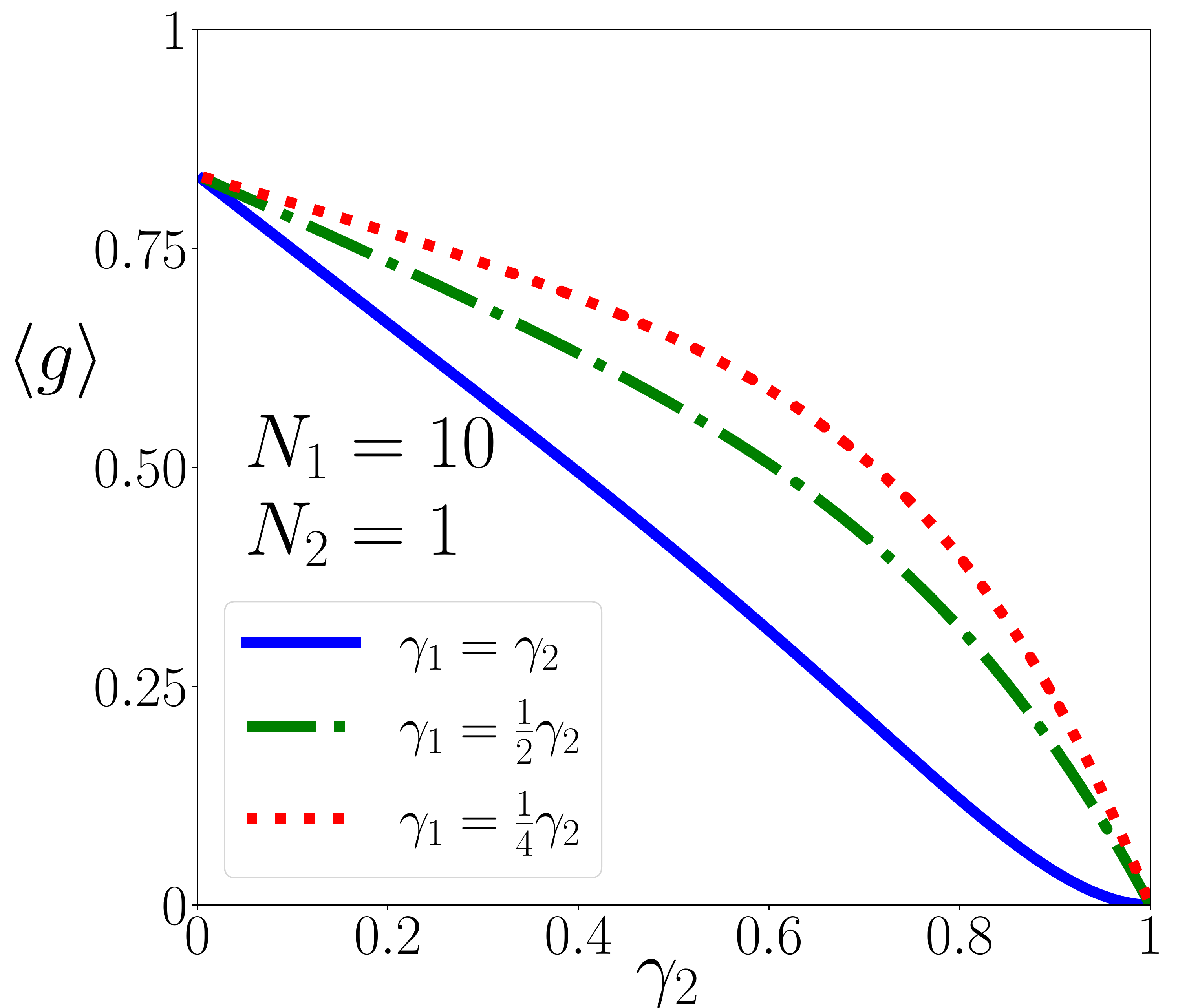}
\caption{(color online) Average conductance for broken time reversal symmetry, as a function of the reflectivities, in several different regimes.}
\end{figure*}

We also treat higher transport moments, but only for broken TRS and $\gamma_2=\gamma_1$. Our calculations suggest that, in this case, some specific quantities (that include all linear statistics $\langle \Tr(t^\dagger t)^n\rangle$, and conductance moments $\langle g^n\rangle$ up to the third), are actually polynomials in $\gamma_1$. This is discussed in Secion V. 

\section{Semiclassical matrix integrals for conductance}

The semiclassical approximation to quantum chaotic transport has been extensively discussed in the past, we refer the reader to previous works \cite{essen3,essen5,greg1,matrix} for details. The matrix element $t_{oi}$ is written in terms of scattering trajectories entering the system through channel $i$ and exiting through channel $o$. When correlations among scattering trajectories are taken into account, and the required integrations over phase space have been performed, the theory has a diagrammatic formulation in terms of ribbon graphs, i.e. a kind of Feynman diagrams. This is a perturbative theory in the parameter $M^{-1}$ which has a topological neature: the contribution of a diagram depends on its Euler characteristic.

Concretely, in the ideal case with no tunnel barriers the computation of a quantity like 
\be\label{moments} \left\langle \prod_{k=1}^n  t^\dagger_{i_ko_{\pi(k)}}t_{o_ki_k}\right\rangle,\ee
for some permutation $\pi$ acting on the outgoing channels labels, requires ribbon graphs with $2n$ vertices of valence $1$ and any number $V$ of vertices of even valence, whose contribution is equal to $(-1)^VM^\chi$, where $\chi=V-E+2n$ is the Euler characteristic, with $E$ being the number of edges. In other words, each edge contributes a factor $M^{-1}$, and each vertex a factor $-M$.

As discussed in \cite{whitney,kuipers}, in the non-ideal case when tunnel barriers are present, the semiclassical diagrammatic rules must be modified. Each edge now contributes \be \widetilde{M}^{-1}=(M-N_1\gamma_1-N_2\gamma_2)^{-1},\ee and each vertex of valence $2q$ contributes
\be-N_1(1-\gamma_1^q)-N_2(1-\gamma_2^q)=-M+N_1\gamma_1^q+N_2\gamma_2^q.\ee Moreover, the final result must be multiplied by $(1-\gamma_1)^n(1-\gamma_2)^n$, to account for transmission through the barriers, and by $\gamma_i$ for each time a trajectory experiences an internal reflection at lead $i$.

Recently \cite{pedro}, two of the present authors combined these diagrammatic rules with the matrix model approach developed in \cite{matrix} to obtain all transport moments as a power series in $\gamma_1$ with coefficients that are rational function of $M$, when $\beta=2$ and $\gamma_2=0$.

The idea of the semiclassical matrix model is to design a integral over matrices with a Gaussian distribution which, by means of Wick's theorem, admit a diagrammatic formulation that coincides with the semiclassical approach to transport moments. 

\subsection{Broken TRS}

When there are no tunnel barriers, the semiclassical rules can be implemented for (\ref{moments}) by means of the matrix integral \cite{matrix}
\be \label{matrix}\lim_{N\to 0}\frac{1}{\mathcal{Z}}\int e^{-M\sum_{q=1}^\infty \frac{1}{q}\Tr(Z^\dag Z)^q} \prod_{k=1}^nZ^\dagger_{i_ko_{\pi(k)}}Z_{o_ki_k}dZ,\ee
where we integrate over $N\times N$ complex matrices $Z$, each matrix element being independently integrated over the complex plane. The quantity
$\mathcal{Z}=\int e^{-M\Tr(Z^\dag Z)}dZ$ is a normalization constant. 

To derive a diagrammatic approach to (\ref{matrix}), $e^{-M\Tr(Z^\dag Z)}$ is kept as a Gaussian measure and the remaining exponentials are expanded as power series. The integral is then performed using Wick's rule, as discussed for example in \cite{zvonkin}. The diagrams thus produced are exactly the semiclassical ones, with the same diagrammatic rules, except for one problem: when producing all possible connections as per Wick's rule, summation over free indices in the traces produces powers of $N$. These are closed cycles that correspond to periodic orbits, forever trapped inside the system. Since the semiclassical approach to transport includes only scattering orbits, we let $N\to 0$ at the end of the calculation to remove these unwanted creatures. 

As discussed in \cite{pedro}, the modified rules that apply to systems with tunnel barriers  can be incorporated, when computing the average conductance, by means of the integral
\be\label{comp} \lim_{N\to 0}(1-\gamma_1)(1-\gamma_2)\frac{1}{\mathcal{Z}}\int \mathcal{I}(Z,Z^\dagger)\mathcal{C}(Z,Z^\dagger)dZ,\ee
where 
\be\mathcal{I}(Z)=e^{-\sum_{q=1}^\infty\frac{1}{q}(M-\gamma_1^qN_1 -\gamma_2^qN_2 )\Tr(Z^\dag Z)^q} \ee
is the internal part, containing all the encounters and segments of trajectories that are inside the chaotic scattering region, and
\be \mathcal{C}(Z,Z^\dagger)= \left(\frac{1}{1-\gamma_2 Z^\dag Z}Z^\dagger\right)_{io}\left(Z\frac{1}{1-\gamma_1 Z^\dag Z}\right)_{oi}\ee
is the channel part, containing information about what happens at the leads. The geometric series, in particular, produce the encounters in which reflections are present. The new normalization constant is $\mathcal{Z}=\int e^{-\widetilde{M}\Tr(Z^\dag Z)}dZ$. 

\subsection{Intact TRS}

A semiclassical matrix model was developed for systems with time-reversal symmetry in \cite{trs}. In particular, the average conductance is given by
\be\label{gtrs}[A_{io}A^*_{io}]\lim_{N\to 0}\frac{1}{\mathcal{Z}}\int e^{-\frac{M}{2}\sum_{q=1}^\infty \frac{1}{q}\Tr(Z^T Z)^q} R_{ii}R_{oo}dZ,\ee
where $Z$ is now real and $Z^T$ is its transpose (here $\mathcal{Z}=\int e^{-\frac{M}{2}\Tr(Z^TZ)}dZ$).

This model is a bit more complicated than for $\beta=2$ because we must make use of the matrix $R=WZW^\dagger$, with $W$ being $M\times N$. The notation $[A_{io}A^*_{io}]$ means we should extract from the result the coefficient of $A_{io}A^*_{io}$, where $A=WW^T$.

In analogy with the previous section, tunnel barriers are included by modifying the internal part of the integrand to be 
\be e^{-\frac{1}{2}\sum_{q=1}^\infty\frac{1}{q}(M-\gamma_1^qN_1 -\gamma_2^qN_2 )\Tr(Z^T Z)^q}\ee
and the channel part to be
\be \left(WZ\frac{1}{1-\gamma_1 Z^TZ}W^\dagger\right)_{ii} \left(W\frac{1}{1-\gamma_2 Z^TZ}Z^TW^\dagger\right)_{oo},\ee
with the new normalization constant being $\mathcal{Z}=\int e^{-\frac{1}{2}\widetilde{M}\Tr(Z^T Z)}dZ$. 

\subsection{Normalization constant}

The normalization constant is  $\mathcal{Z}=\int e^{-\frac{\beta}{2}\widetilde{M}\Tr(Z^\dagger Z)}dZ,$ with $\beta=2$ for broken TRS (complex $Z$) and $\beta=1$ for intact TRS (real $Z$). In both cases we use the singular value decomposition $Z=UDV^\dagger$, with $U$ and $V$ random matrices from the unitary/orthogonal group for $\beta=2/1$. Let $\int dUdV=\mathcal{G}$. 

The Jacobian of the singular value decomposition is $|\Delta(X)|^{\beta}\det(X)^{\frac{\beta}{2}-1}$ \cite{jacob,forresterbook}, where $X=D^2$ is a diagonal matrix with the same eigenvalues, $\{x_1,...,x_N\}$, of $Z^\dagger Z$ and 
\be \Delta(X)=\prod_{1\le i<j\le N}(x_j-x_i) \ee
is the so-called Vandermonde. This leads to 
\be \mathcal{Z}=\mathcal{G}\int_0^\infty e^{-\frac{\beta}{2}\widetilde{M} {\rm Tr}(X)}|\Delta(X)|^{\beta}\det(X)^{\frac{\beta}{2}-1}dX.\ee

In what follows we shall use for convenience the Jack parameter
\be \alpha=\frac{2}{\beta},\ee
instead of the Dyson parameter $\beta$. Then
\be \mathcal{Z}_\alpha=\mathcal{G}_\alpha\int_0^\infty e^{-\frac{\widetilde{M}}{\alpha} {\rm Tr}(X)}|\Delta(X)|^{2/\alpha}\det(X)^{\frac{1}{\alpha}-1}dX,\ee
which is a Selberg integral \cite{selberg}, known to be equal to
\be \mathcal{Z}_\alpha=\mathcal{G}_\alpha \left(\frac{\alpha}{\widetilde{M}}\right)^{N^2/\alpha}\prod_{j=1}^{N}\frac{\Gamma(j/\alpha)\Gamma(1+j/\alpha)}{\Gamma(1+1/\alpha)}.\ee

\section{Solution for conductance}

\subsection{Angular integrals}

Both versions of the matrix integral are computed by means of the singular value decomposition $Z=UDV^\dagger$. The internal part is independent of $U,V$. For broken TRS, the channel part is 
\be \sum_{a,b=1}^N V_{ia}\frac{D_a}{1-\gamma_2 X_a}U^\dagger_{ao}U_{ob}\frac{D_b}{1-\gamma_1 X_b}V^\dagger_{bi},\ee
while for intact TRS it is 
\be \sum_{c,d,e,f,g,h=1}^N W_{ic}\frac{U_{cd}D_dV^T_{de}}{1-\gamma_1 X_d}W^\dagger_{ei}W_{of}\frac{V_{fg}D_gU^T_{gh}}{1-\gamma_2 X_g}W^\dagger_{ho}.\ee

The integrals require only the orthogonality relations of matrix elements of the classical compact Lie groups,
\be\frac{1}{\mathcal{G}_{\alpha}}\int U_{ab}U^\dagger_{cd}\int V_{ef}V^\dagger_{gh}dUdV=\frac{\delta_{ad}\delta_{bc}\delta_{eh}\delta_{fg}}{N^2}.\ee 

For broken TRS, the result of the angular integral is
\begin{multline} \frac{1}{N^2}\sum_{c=1}^N \frac{D^\dagger_c}{1-\gamma_2 X_c}\frac{D_c}{1-\gamma_1 X_c} \\= \frac{1}{N^2} {\rm Tr}\left(\frac{X}{(1-\gamma_1 X)(1-\gamma_2 X)}\right).\end{multline}
For intact TRS, we have
\be \frac{1}{N^2}\sum_{c,d,e=1}^N W_{ic}W^\dagger_{ei}W_{oe}W^\dagger_{co}\frac{X_d}{(1-\gamma_1 X_d)(1-\gamma_2 X_d)}\ee
or 
\be A_{io}A^*_{io}\frac{1}{N^2}{\rm Tr}\left(\frac{X}{(1-\gamma_1 X)(1-\gamma_2 X)}\right).\ee
After we collect the coefficient of $A_{io}A^*_{io}$, as required by the theory (c.f. Eq.(\ref{gtrs})), we arrive at the same result as for broken TRS.

In both cases, the sums over $i,j$ produce a factor $N_1N_2$ which, combined with the prefactors $(1-\gamma_1)(1-\gamma_2)$ already present, becomes $\tn_1\tn_2$.

\subsection{Eigenvalue integrals}

Having computed the angular integrals, we define 
\be \mathcal{C}(X)={\rm Tr}\left(\frac{X}{(1-\gamma_1 X)(1-\gamma_2 X)}\right).\ee
and are left to deal with the eigenvalue integral
\be\label{E} E=\int _0^1|\Delta(X)|^{2/\alpha}\det(X)^{\frac{1}{\alpha}-1}
\mathcal{I}(X)\mathcal{C}(X)dX,\ee
with the internal part $\mathcal{I}(X)$ now written as
\be \det(1-X)^{\frac{M}{\alpha}}\det(1-\gamma_1 X)^{-\frac{N_1}{\alpha}}\det(1-\gamma_2 X)^{-\frac{N_2}{\alpha}},\ee
where we have used 
\be e^{-N\sum_{q=1}^\infty \frac{1}{q} {\rm Tr}X^q}=e^{N{\rm Tr}\log(1-X)}=\det(1-X)^{N}.\ee

In order to compute the eigenvalue integral, we resort to the theory of Jack polynomials $J_\lambda^{(\alpha)}(x_1,...,x_N)$, labelled by an integer partition $\lambda=(\lambda_1,...,\lambda_\ell)$ with number of parts less or equal to $N$, $\ell(\lambda)\le N$. When $\alpha=1$ they are proportional to Schur polynomials, when $\alpha=2$ they are called zonal polynomials.

Jack polynomials are useful in this context because they satisfy the beautiful Jack-Selberg integral \cite{kaneko,kadell}, which states that
\be \int_0^1 |\Delta(X)|^{2/\alpha}\det(X)^{p-1}\det(1-X)^{q-1}J_\lambda^{(\alpha)}(X)dX\ee
is equal to 
\begin{multline} J_\lambda^{(\alpha)}(1^N)\prod_{j=1}^{N}\frac{\Gamma(1+j/\alpha)}{\Gamma(1+1/\alpha)}\\\times\prod_{j=1}^{N}\frac{\Gamma(\lambda_j+p+(N-j)/\alpha)\Gamma(q+(N-j)/\alpha)}{\Gamma(\lambda_j+p+q+(2N-j-1)/\alpha)}.\end{multline}

When their argument is the identity matrix, we have
\be J_\lambda^{(\alpha)}(1^N)=\alpha^{|\lambda|}\prod_{j=1}^{N}\frac{\Gamma(\lambda_j+(N-j+1)/\alpha)}{\Gamma((N-j+1)/\alpha)}.\ee We use a special notation for this quantity:
\be [N]_\lambda^{(\alpha)}=J_\lambda^{(\alpha)}(1^N).\ee

In the following calculations we rely on some well known results about Jack polynomials, which are collected in the Appendix for convenience.

\subsection{Broken TRS}

In order to be able to employ the Jack-Selberg integral, we must expand the interior part of the integrand as an infinite series, using the Cauchy expansion
\be \det(1-XY)^{-1}=\sum_{\lambda}s_\lambda(X)s_\lambda(Y).\ee
Notice that $\det(1-\gamma_i X)^{-N_i/\alpha}=\det(1-\hat{\gamma_i} X)^{-1/\alpha}$, where $\hat{\gamma_1}$ is a matrix given by $\gamma_i$ times the $N_i$-dimensional identity. If we define a combined matrix $\hat{\gamma_c}$ of dimension $M$,
\be \hat{\gamma_c}=\hat{\gamma_1}\oplus\hat{\gamma_2},\ee
then
\be \det(1-\gamma_1 X)^{-N_1}\det(1-\gamma_2 X)^{-N_2}=\sum_{\lambda}s_\lambda(X)s_\lambda(\hat{\gamma_c}).\ee

For the channel part of the integrand we have,
\be \mathcal{C}(X)=\sum_{n,m\ge 0}\gamma_1^n\gamma_2^m{\rm Tr}(X^{n+m+1}),\ee
which can also be expanded in terms of Schur polynomials,
\be \mathcal{C}(X)=\sum_{n,m\ge 0}\gamma_1^n\gamma_2^m\sum_{\mu\vdash n+m+1}\chi_\mu(n+m+1) s_\mu(X),\ee
where $\chi_\mu(\lambda)$ are the characters of the irreducible representations of the permutation group \cite{macdonald}. 

It is well known that $\chi_\mu(n+m+1)=0$ unless $\mu$ is a so-called hook partition, $\mu=(n+m+1-k,1^k)$, in which case it equals $(-1)^k$. When $\lambda\vdash n$, the relation between Schur and Jack polynomials is $s_\lambda=\frac{d_\lambda}{n!}J_\lambda^{(1)}$, where $d_\lambda=\chi_\lambda(1^n)$ is the dimension of the corresponding irreducible representation.

We therefore have two Schur polynomials expansions in our integrand, one coming from the internal part and one from the channel part. We must therefore bring into play the Littlewood-Richardson coefficients, defined as 
\be\label{LR1} s_\lambda(X) s_\mu(X)=\sum_{\theta\vdash |\lambda|+|\mu|} C^{(1)}_{\lambda,\mu,\theta}s_\theta(X).\ee

We find that
\begin{multline}\langle g\rangle=\frac{\tn_1\tn_2}{N^2}\sum_{n,m\ge 0}\gamma_1^n\gamma_2^m\sum_{\mu\vdash n+m+1}\chi_\mu(n+m+1)\\\times\sum_{\theta}s_{\theta/\mu}(\hat{\gamma_c})B_\theta^{(1)},\end{multline}
where we have used the skew polynomials \cite{macdonald}
\be \sum_{\lambda}C^{(1)}_{\lambda,\mu,\theta}s_\lambda(\hat{\gamma_c})=s_{\theta/\mu}(\hat{\gamma_c}),\ee
and the integral that must be done is
\be B_\theta^{(1)}=\frac{\mathcal{G}_1}{\mathcal{Z}_1}\int _0^1|\Delta(X)|^{2}\det(1-X)^{M}s_\theta(X)dX,\ee
which is given by \cite{selberg}
\be \widetilde{M}^{N^2}\left([N]_\theta^{(1)}\right)^2\frac{d_\theta}{|\theta|!}\prod_{j=1}^N\frac{(M+N-j)!}{(\theta_j+M+2N-j)!}.\ee

\subsection{Intact TRS}

The zonal Cauchy expansion is \cite{macdonald}
\be \det(1-XY)^{-1/2}=\sum_{\lambda}\frac{d_{2\lambda}}{(2|\lambda|)!} z_\lambda(X)z_\lambda(Y),\ee where $z_\lambda=J_\lambda^{(2)}$, so
\begin{multline} \det(1-\gamma_1 X)^{-N_1/2}\det(1-\gamma_2 X)^{-N_2/2}\\=\sum_{\lambda}\frac{d_{2\lambda}}{(2|\lambda|)!}z_\lambda(X)z_\lambda(\hat{\gamma_c}).\end{multline}

For the channel part of the integrand we have \cite{macdonald}
\be \mathcal{C}(X)=\sum_{n,m\ge 0}\frac{\gamma_1^n\gamma_2^m}{(2n+2m+1)!}\sum_{\mu\vdash n+m+1}t_2(\mu)d_{2\mu}z_\mu(X),\ee
where \be t_\alpha(\mu)=\prod_{i=1}^{\ell(\mu)}\prod_{j=1}^{\mu_i}[\alpha(j-1)-i+1]\ee
is the product of all non-zero $2$-contents of $\mu$.

Making use of the zonal Littlewood-Richardson coefficients,
\be\label{LR2} z_\lambda(X) z_\mu(X)=\sum_{\theta\vdash |\lambda|+|\mu|} C^{(2)}_{\lambda,\mu,\theta}z_\theta(X),\ee
we end up with
\begin{multline}\langle g\rangle=\frac{\tn_1\tn_2}{N^2}\sum_{n,m\ge 0}\frac{\gamma_1^n\gamma_2^m}{(2n+2m+1)!}\sum_{\mu\vdash n+m+1}d_{2\mu}t_2(\mu)\\\times\sum_{\theta}z_{\theta/\mu}(\hat{\gamma_1},\hat{\gamma_2})B_\theta^{(2)},\end{multline}
where we have used \cite{macdonald} \be \sum_{\lambda}C^{(2)}_{\lambda,\mu,\theta}\frac{d_{2\lambda}}{(2|\lambda|)!}z_\lambda(\hat{\gamma_c})=\frac{d_{2\theta}}{(2|\theta|)!}z_{\theta/\mu}(\hat{\gamma_c}),\ee
and the integral that must be done is
\be B_\theta^{(2)}=\frac{d_{2\theta}}{(2|\theta|)!}\frac{\mathcal{G}_2}{\mathcal{Z}_2}\int _0^1\frac{|\Delta(X)|}{\sqrt{\det(X)}}\det(1-X)^{M/2}z_\theta(X)dX,\ee
which is given by \cite{selberg}
\begin{multline} \left(\frac{\widetilde{M}}{2}\right)^{N^2/2}\left([N]_\theta^{(2)}\right)^2\frac{d_{2\theta}}{2^{|\theta|}(2|\theta|)!}\\\times\prod_{j=1}^N\frac{\Gamma(1+(M+N-j)/2)}{\Gamma(1+\theta_j+(M+2N-j)/2)}.\end{multline}

\subsection{Taking $N\to 0$}

Having computed all the integrals, we must consider the limit $N\to 0$, as discussed in Section III.A. First of all, 
$ \left(\frac{\widetilde{M}}{\alpha}\right)^{N^2/\alpha}\to 1.$
Also,
\be \prod_{j=1}^N\frac{\Gamma(1+(M+N-j)/\alpha)}{\Gamma(1+\theta_j+(M+2N-j)/\alpha)}\to\frac{\alpha^{|\theta|}}{[M+\alpha-1]_\theta^{(\alpha)}}.\ee
Finally, we need to deal with 
\be
 \lim_{N\to 0} \frac{\left([N]_\theta^{(\alpha)}\right)^2}{N^2}.
\ee
From the expression of $[N]_\theta^{(\alpha)}$ in terms of contents, Eq. ({\ref{Nc}), we know that, for small $N$,
\be [N]_\theta^{(\alpha)}= t_\alpha(\theta) N^{D_\alpha(\theta)}+O(N^{D_\alpha(\theta)+1}),\ee 
where $t_\alpha$ and $ D_\alpha$ are discussed in the Appendix. Therefore, we must have $D_\alpha(\theta)=1$ and
\be
\lim_{N\to 0} \frac{\left([N]_\theta^{(\alpha)}\right)^2}{N^2}=(t_\alpha(\theta))^2.
\ee

We thus have very similar results for the two symmetry classes. For broken TRS,
\begin{widetext}
\be\label{g1}\langle g\rangle=\tn_1\tn_2\sum_{n,m\ge 0}\gamma_1^n\gamma_2^m\sum_{\mu\vdash n+m+1}(-1)^{\ell(\mu)-1}\sum_{\theta, D_1(\theta)=1}s_{\theta/\mu}(\hat{\gamma_c})\frac{d_\theta}{|\theta|!}\frac{(t_1(\theta))^2}{[M]_\theta^{(1)}}.\ee For intact TRS,
\be\label{g2}\langle g\rangle=\tn_1\tn_2\sum_{n,m\ge 0}\gamma_1^n\gamma_2^m\sum_{\mu\vdash n+m+1}\frac{d_{2\mu}t_2(\mu)}{(2n+2m+1)!}\sum_{\theta, D_2(\theta)=1}z_{\theta/\mu}(\hat{\gamma_c})\frac{d_{2\theta}}{(2|\theta|)!}\frac{(t_2(\theta))^2}{[M+1]_\theta^{(2)}}.\ee
\end{widetext}

\subsection{Explicit expression for $\beta=2$}

When $D_1(\theta)=1$, $\theta$ is a hook partition. In that case, we have 
\be \frac{d_\theta(t_1(\theta))^2}{|\theta|!}=\frac{|t_1(\theta)|}{|\theta|}.\ee 
If we write $\theta=(n+m+1-k+r,1^{k+q})$, then 
\be |t_1(\theta)|= (n+m-k+r)!(k+q)!,\ee
and
\be[M]^{(1)}_{\theta} = M(M+1)^{(n+m-k+r)}(M-1)_{(k+q)}.\ee

Moreover, because $\mu$ is also a hook partition, say $\mu=(n+m+1-k,1^k)$, the diagram of $\theta/\mu$ has two disjoint pieces, $\theta/\mu=(r)\cup (1^q)$. Then the skew polynomial factors, $s_{\theta/\mu}(\hat{\gamma}_c) = s_{(r)}(\hat{\gamma}_c)s_{(1^q)}(\hat{\gamma}_c)$ (unfortunately, $z_{\theta/\mu}(\hat{\gamma_c})$ does not factor so nicely). 

But $s_{(r)} = h_r$ is the complete symmetric polynomial, while $s_{(1^q)}= e_q$ is the elementary symmetric polynomial. Using the generating function for these polynomials, 
\begin{align}
\sum_{r=0}^{\infty} h_r(\hat{\gamma}_1,\hat{\gamma}_2) t^r = \frac{1}{(1-\gamma_1 t)^{N_1}}\frac{1}{(1-\gamma_2 t)^{N_2}}\\
\sum_{q=0}^{M} e_q(\hat{\gamma}_1,\hat{\gamma}_2) t^q = (1+\gamma_1 t)^{N_1}(1+\gamma_2 t)^{N_2}
\end{align}
and expanding the right hand side of both expressions, we get
\begin{align}
h_r(\hat{\gamma}_1,\hat{\gamma}_2) &= \sum_{i= 0}^r \frac{(N_1)^{(i)}(N_2)^{(r-i)}}{i!(r-i)!}\gamma_1^i \gamma_2^{r-i},\\
e_q(\hat{\gamma}_1,\hat{\gamma}_2) &= \sum_{l= 0}^q \frac{(N_1)_{(l)}(N_2)_{(q-l)}}{l!(q-l)!}\gamma_1^l \gamma_2^{q-l}.
\end{align}

In the end,
\be
   \langle g\rangle = \frac{\tn_1\tn_2}{M} \sum_{n,m,r,q\ge 0}\sum_{k=0}^{n+m} \sum_{i= 0}^r\sum_{l=0}^qG^{n,m,k}_{r,q,i,l} \gamma_1^{l+i+n} \gamma_2^{r+q+m-l-i}
\ee
where
\begin{multline}
    G^{n,m,k}_{r,q,i,l} = \frac{(-1)^k(n+m-k+r)!(k+q)!}{i!(r-i)!l!(q-l)!(n+m+r+q+1)}\\\times\frac{(N_1)^{(i)}(N_2)^{(r-i)}}{(M+1)^{(n+m-k+r)}} \frac{(N_1)_{(l)}(N_2)_{(q-l)}}{(M-1)_{(k+q)}}.
\end{multline}

When $\gamma_2=\gamma_1$, the expression can be simplified to
\begin{multline}
   \langle g\rangle = \tn_1\tn_2M \sum_{m,r,q\ge 0}\sum_{k=0}^{m}\frac{(-1)^k(m-k+r)!(k+q)!}{r!q!(m+r+q+1)}\\\times\frac{(m+1)\gamma_1^{m+r+q}}{(M+r)^{(m-k+1)}(M-q)_{(k+1)}}.
\end{multline}
When the terms in this series are actually computed it seems almost all of them vanish, except three. We cannot prove this rigorously, but have verified it extensively. The two lowest order terms are easy to compute, they are given by $\frac{N_1N_2}{M}$ and $-\frac{N_1N_2M\gamma_1}{M^2-1}$. The only other nonvanishing term is of order $\gamma_1^M$, and its computation requires some extra care. We must make $M=N_1+N_2+\epsilon$ and delay the limit $\epsilon\to 0$ until after all cancellations, because in general the term of order $\gamma_1^t$ is given by
\be \frac{2N_1N_2\epsilon\gamma_1^t}{(M^2-1)(M+\epsilon+t)(M+\epsilon-t)}.\ee
This vanishes when $\epsilon\to 0$, unless $t=M$. The final result is 
\be\frac{N_1N_2}{M}-\frac{N_1N_2M\gamma_1}{M^2-1}+\frac{N_1N_2\gamma_1^M}{M(M^2-1)}.\ee In particular, when $N_1=N_2=1$ we get $1/2-2\gamma_1/3+\gamma_1^2/6$, in agreement with the corresponding random matrix theory result \cite{BB2}, but in disagreement with a previous perturbative semiclassical result \cite{kuipers}, which is correct up to the first few orders in $1/M$ but was not able to capture the non-perturbative term $\gamma_1^M$.

\subsection{Generalization to many leads}

If the system has $L$ leads, each with its own tunnel barrier of transparency $\Gamma_j=1-\gamma_j$, the theory can be generalized directly. All that is required to compute the conductance between a pair of leads is to build exactly the same matrix integrals, except that the internal term should be
\be \mathcal{I}(Z)=e^{-\frac{1}{\alpha}\sum_{q=1}^\infty \frac{1}{q}(M-\sum_{j=1}^L\gamma_j^qN_j)}.\ee Likewise, when using the Cauchy expansion one should define the combined matrix
\be \hat{\gamma}_c=\bigoplus_{j=1}^L \hat{\gamma}_j.\ee
Eqs. (\ref{g1}) and (\ref{g2}) remain valid, with $M=\sum_{j=1}^L N_j$.

\section{Higher moments}

The present approach is also able to treat transport moments higher than conductance, but only if we assume broken TRS and the transparency of both leads to be identical, $\gamma_2=\gamma_1$. 

Schur functions are a basis for the vector space of homogeneous symmetric polynomials, so transport moments can be written as linear combinations of them. In fact,
\be p_\mu(t^\dagger t)=\sum_{\lambda} \chi_\lambda(\mu)s_\lambda(t^\dagger t),\ee
where 
\be p_\mu(X)=\prod_{i=1}^{\ell(\mu)}\Tr(X^{\mu_i})\ee
and $\chi_\lambda(\mu)$ are irreducible characters of the permutation group.

As discussed in \cite{pedro}, when computing $\langle s_\lambda(t^\dagger t)\rangle$, the channel factor is 
\be s_\lambda\left(\frac{1}{1-\gamma Z^\dagger Z}Z^\dagger Q_1 Z \frac{1}{1-\gamma Z^\dagger Z} Q_2\right),\ee where $Q_i$ is diagonal with $N_i$ eigenvalues equal to $1$ and the rest equal to zero. The angular integrals are
\be \int dU dV s_\lambda\left(V\frac{1}{1-\gamma X}DU^\dagger Q_1 UD \frac{1}{1-\gamma X}V^\dagger Q_2\right).\ee
Using that
\be \int dU s_\lambda(UAU^\dagger B)=\frac{s_\lambda(A)s_\lambda(B)}{s_\lambda(1^N)},\ee
they can be carried out to produce
\be \frac{[N_1]_\lambda[N_2]_\lambda}{[N]_\lambda^2}f_\lambda(X),\ee where
\be f_\lambda(X)=s_\lambda\left(\frac{X}{(1-\gamma X)^2}\right).\ee On the other hand, the eigenvalue integral is
\be \frac{1}{\mathcal{Z}}\int \frac{\det(1-X)^M}{\det(1-\gamma X)^{M}}|\Delta(X)|^2f_\lambda(X)dX.\ee

\subsection{Expanding $f_\lambda(X)$}

We must write $f_\lambda(X)$ as a linear combination of Schur polynomials of $X$. Let us start by using a ratio of determinants,
\be s_\lambda(X)=\frac{\det\left(x_j^{N+\lambda_i-i}\right)}{\det\left(x_j^{N-i}\right)}=\frac{\det\left(x_j^{N+\lambda_i-i}\right)}{\Delta(X)}.\ee
In the present case this gives
\be f_\lambda(X)=\det\left[\left(\frac{x_k}{(1-\gamma x_k)^2}\right)^{N+\lambda_i-i}\right]\frac{1}{\Delta\left(\frac{X}{(1-\gamma X)^2}\right)}.\ee
It is easy to express the above Vandermonde as
\be \Delta\left(\frac{X}{(1-\gamma X)^2}\right)=\frac{\Delta(X)\prod_{i<j}(1-\gamma^2 x_ix_j)}{\det(1-\gamma X)^{2N-2}}.\ee
Now, we make use of the Littlewood identity \cite{macdonald}
\be \frac{1}{\prod_{i<j}(1-\gamma^2 x_ix_j)}=\det(1-\gamma X)\sum_\mu s_\mu(\gamma X),\ee
to arrive at
\be f_\lambda(X)=\det\left[\left(\frac{x_k^{N+\lambda_i-i}}{(1-\gamma x_k)^{2\lambda_i-2i+1}}\right)\right]\frac{1}{\Delta(X)}\sum_\mu s_\mu(\gamma X).\ee

We still must expand
\be \det\left[\left(\frac{x_k^{N+\lambda_i-i}}{(1-\gamma x_k)^{2\lambda_i-2i+1}}\right)\right]\frac{1}{\Delta(X)}\ee as a combination of Schur polynomials. This is done by using their orthogonality when integrated around the unit circle in the complex plane \cite{meckes},
\be \frac{1}{N!}\oint s_\lambda(z)s_\rho(\bar{z})|\Delta(z)|^2=\delta_{\lambda\rho}.\ee
The expansion coefficients we are looking for are thus
\be \frac{1}{N!}\oint \det\left[\left(\frac{z_k^{N+\lambda_i-i}}{(1-\gamma z_k)^{2\lambda_i-2i+1}}\right)\right]s_\rho(\bar{z})\Delta(\bar{z}).\ee
This can be computed by means of the Andreief identity,
\be \int \det(f_i(x_k))\det(g_j(x_k))=N!\det\left[\int f_i(x)g_j(x)\right],\ee
which leads to 
\be \det\left[\oint \frac{z^{N+\lambda_i-i}\bar{z}^{N+\rho_j-j}}{(1-\gamma z_k)^{2\lambda_i-2i+1}}\right].\ee
Since $z^N\bar{z}^N=1$, the binomial theorem gives
\be \det\left[ \sum_{k=0}^\infty \binom{2\lambda_i-2i+k}{2\lambda_i-2i}\gamma^k\oint z^{\lambda_i-i+k}\bar{z}^{\rho_j-j}\right].\ee
The integral is only different from zero if $k=\rho_j-j-\lambda_i+i$. Since $k\ge 0$ we have that $\rho_j-j\ge\lambda_i-i$ and it reduces to
\be \gamma^{|\rho|-|\lambda|}\det\left[ \binom{\rho_j+\lambda_i-i-j}{2\lambda_i-2i}\right].\ee

So we found that
\be f_\lambda(X)=\sum_\rho \gamma^{|\rho|-|\lambda|}R_{\lambda,\rho}s_\rho(X)\sum_\mu s_\mu(\gamma X),\ee or 
\be f_\lambda(X)=\sum_\rho \gamma^{|\rho|-|\lambda|+|\mu|}R_{\lambda,\rho}\sum_\mu \sum_\nu C^{(1)}_{\rho\mu\nu} s_\nu(X),\ee
where $C^{(1)}$ are the Littlewood-Richardson coefficients and
\be R_{\lambda,\rho}=\det\left[ \binom{\rho_j+\lambda_i-i-j}{2\lambda_i-2i}\right].\ee

The integral to be done is thus 
\be \frac{1}{\mathcal{Z}}\int \det(1-X)^M\det(1-\gamma X)^{-M}|\Delta(X)|^2s_\nu(X).\ee
We use again the Cauchy expansion and write the result in terms of a skew Schur polynomial,
\begin{multline} \langle s_\lambda(t^\dagger t)\rangle=(1-\gamma)^{2n}[N_1]_\lambda[N_2]_\lambda\\\times\sum_{\underset{D(\theta)=D(\lambda)}{\rho,\mu,\nu,\theta}} \gamma^{|\theta|-|\lambda|}R_{\lambda,\rho}s_{\theta/\nu}(1^{M})C^{(1)}_{\rho\mu\nu}\frac{d_\theta t_1(\theta/\lambda)^2}{|\theta|![M]_\theta}.\end{multline}
The term $\gamma^0$ comes from $\rho=\lambda=\nu=\theta$ and $\mu=\emptyset$, so we have the correct result,
\be \langle s_\lambda(t^\dagger t)\rangle=\frac{[N_1]_\lambda[N_2]_\lambda}{[M]_\lambda}\frac{d_\lambda}{n!}+O(\gamma).\ee 

\subsection{Hook moments}

If $\lambda$ is a hook, i.e. a partition of the form $(n-k,1^k)$, we call $\langle s_\lambda(t^\dagger t)\rangle$ a hook moment. Transport moments that can be calculated from hook moments include all linear statistics, because $\chi_\lambda(n)$ is different from zero only if $\lambda$ is a hook. Besides, all partitions of $n$ are hooks if $n\le 3$. 

In this case $Q_{\lambda,\rho}$ is different from zero only if $\rho$ is also a hook. Moreover, $\theta$ must be a hook and then $s_{\theta/\nu}(1^M)$ is different from zero only if $\nu$ is also a hook. Finally, $\rho$ and $\nu$ being hooks imply that $\mu$ is a hook. Then we can use
\be R_{(A,1^a),(B,1^b)}=(-1)^{a+b}\binom{B+A-2}{2A-2}\binom{b+a}{2a}\frac{2b+1}{2a+1}.\ee
and
\be s_{(T,1^t)/(V,1^v)}(1^{M})=\frac{(M)^{(T-V)}(M)_{(t-v)}}{(T-V)!(t-v)!}.\ee
to get manageable expressions.

Extensive computer calculations suggest that, when $\lambda=(n-k,1^k)$, the moment $\langle s_\lambda(t^\dagger t)\rangle$ is actually a polynomial in $\gamma$ and that this polynomial has two blocks of terms: the first block contains elements of order $\gamma^m$ with $0\le m\le n$, while the other block contains elements of order $\gamma^m$ with $M-k\le m\le M+n-k-1$ (remember that $[N_1]_\lambda$ vanishes unless $N_1\ge\ell(\lambda)=k+1$, so we must restrict $M\ge 2k+2$). That is,
\be\label{hooks} \langle s_{(n-k,1^k)}(t^\dagger t)\rangle=P_\lambda(\gamma)+\gamma^{M-k}Q_\lambda(\gamma),\ee
where $P_\lambda(\gamma)$ is a polynomial of degree $n$ and $Q_\lambda(\gamma)$ is a polynomial of degree $n-1$. The calculation of this second family of polynomials is delicate: just like we did in Section IV.F, it is necessary to set $M=N_1+N_2+\epsilon$, take into account all simplifications, and then let $\epsilon\to 0$.

Why the expressions (\ref{hooks}) should be so simple is a mistery. Notice that the theory contains a prefactor $(1-\gamma)^{2n}$ coming from the entry and exit of trajectories, and an infinite series in $\gamma$ coming from the encounters taking place in the chaotic region. Somehow, cancellations lead to a polynomial. 

Let us remark that, just like for conductance, the second polynomial in (\ref{hooks}) is not algebraic in $M$, being exponentially small for large $M$ and therefore not accessible to semiclassical methods that only give the leading orders in a $1/M$ expansion. 

For example, for conductance we have $n=1$, $k=0$. In that case, we believe (but cannot prove) that
\begin{align} \langle g\rangle&=\tn_1\tn_2\sum_{\underset{D(\theta)=1}{\rho,\mu,\nu,\theta}} \gamma^{|\theta|-1}R_{(1),\rho}s_{\theta/\nu}(1^{M})c^\nu_{\rho\mu}\frac{|t_1(\theta)|}{|\theta|[M]_\theta}\\&=\tn_1\tn_2\sum_{q=0}^\infty\frac{\gamma^q(M^2-q-1)}{M(M^2-1)}+\frac{N_1N_2\gamma^M}{M(M^2-1)}\\&=\frac{N_1N_2}{M}\left(1-\frac{M^2\gamma}{M^2-1}\right)+\frac{N_1N_2\gamma^M}{M(M^2-1)}.\end{align}
In other words, in this case we have
\be P_{(1)}(\gamma)=\frac{N_1N_2}{M}-\frac{N_1N_2M\gamma}{M^2-1},\ee
and
\be Q_{(1)}(\gamma)=\frac{N_1N_2}{M(M^2-1)}.\ee

Moments of order $2$ are required in order to compute important statistics like conductance variance and average shot-noise. Concretely,
\be {\rm var}(g)=\langle s_{(1,1)}(t^\dagger t)\rangle+\langle s_{(2)}(t^\dagger t)\rangle-\langle g\rangle^2,\ee
and
\be \Tr[t^\dagger t(1-t^\dagger t)]=g+s_{(1,1)}(t^\dagger t)-s_{(2)}(t^\dagger t).\ee
What we find is that
\begin{widetext}
\be \langle P_{(2)}(\gamma)\rangle=(N_1)^{(2)}(N_2)^{(2)}\left(\frac{1}{2M(M+1)}-\frac{M\gamma}{(M-1)(M+1)(M+2)}+\frac{(M+1)\gamma^2}{2(M-1)(M+2)(M+3)}\right),
\ee
and
\be \langle P_{(1,1)}(\gamma)\rangle=(N_1)_{(2)}(N_2)_{(2)}\left(\frac{1}{2M(M-1)}-\frac{M\gamma}{(M+1)(M-1)(M-2)}+\frac{(M-1)\gamma^2}{2(M+1)(M-2)(M-3)}\right).
\ee
\end{widetext}

The polynomials $Q_{(2)}$ and $Q_{(1,1)}$ are more complicated and we could not find a  general formula for them. We just present some special cases with small channel numbers. When $N_1+N_2=3$ (one is $1$ and the other is $2$),
\be Q_{(2)}=\frac{1}{10}(1-\gamma).\ee
When $N_1+N_2=4$ (either both are $2$ or one is $1$ and the other is $3$),
\be Q_{(2)}=(N_1)^{(2)}(N_2)^{(2)}\left(\frac{1}{360}-\frac{\gamma}{315}\right).\ee
When $N_1=N_2=2$,
\be Q_{(1,1)}=\frac{4}{15}\left(-1+\frac{\gamma}{8}\right).\ee

\section{Conclusion}

By using a formulation in terms of matrix integrals, we developed a semiclassical approach to quantum chaotic transport that is able to describe systems with tunnel barriers in the leads. Our results incorporate the barriers in a perturbative way, as power series in their reflectivities, but are exact in the number of channels, i.e. there is no large-$M$ expansion. We obtained new expressions for the average conductance, both for systems with and without time-reversal symmetry. We also obtained higher order moments, like conductance variance and shot-noise, when time-reversal is broken and the two leads are identical. In particular, our method is able to obtain non-perturbative contributions like $\gamma^M$, which were not accessible to previous semiclassical approaches which were restricted to leading orders in $1/M$.

\section*{Acknowledgments}

Financial support from CAPES and from CNPq, grant 306765/2018-7, are gratefully acknowledged. We have profited from discussions with Jack Kuipers.

\section*{Appendix}

A partition $\lambda\vdash n$ can be represented by a diagram, which is a left-justified collection of boxes containing $\lambda_i$ boxes in line $i$. In Figure \ref{diagram} we show the diagram associated with $\lambda=(4,4,2,2,1)$. The $j$th box in line $i$ is denoted by $(i,j)$ and its $\alpha$-content is given by 
\be c_\alpha(j,i)=\alpha(j-1)-i+1.\ee These contents are also shown in Figure \ref{diagram}. 

The Durfee $\alpha$-rectangle of $\lambda$ is the largest rectangle covered by the diagram of $\lambda$ whose lower-right corner has zero $\alpha$-content. Let $D_\alpha(\lambda)$ be the horizontal size of the Durfee $\alpha$-rectangle of $\lambda$, i.e., the number of boxes in $\lambda$ with zero $\alpha$-content. These rectangles are  highlighted in grey in Figure \ref{diagram}.

\begin{figure*}[t]
\includegraphics[scale=0.7,clip]{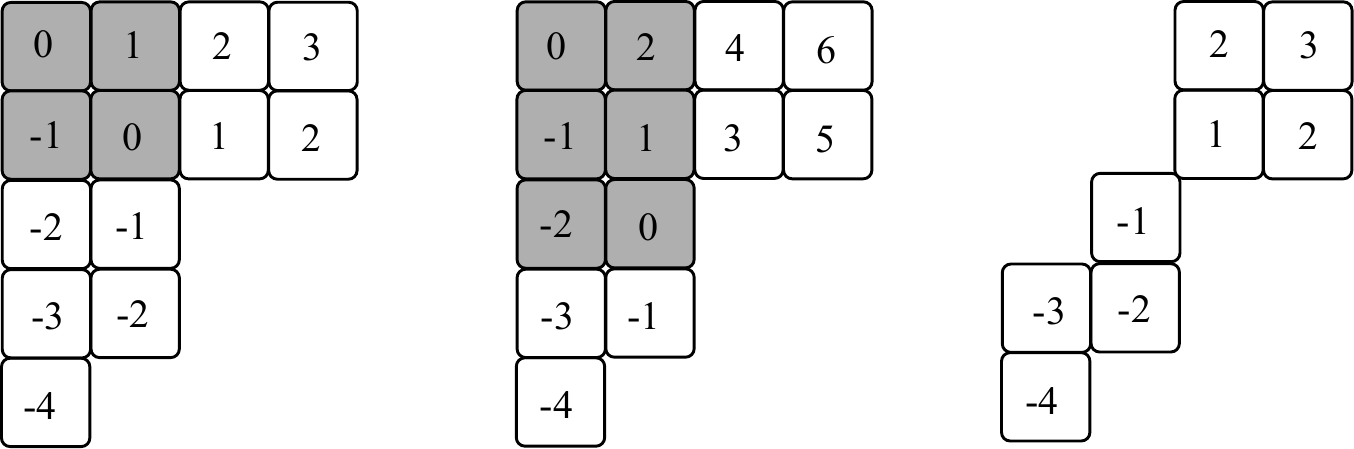}
\caption{Left: Diagram of the partition $(4,4,2,2,1)$, showing $1$-contents; the Durfee square is highlighted in grey. Middle: Diagram of the same partition, but showing $2$-contents; the Durfee $2$-rectangle is highlighted in grey. Right: the skew diagram $(4,4,2,2,1)/(2,2,1)$ and its $1$-contents.} 
\label{diagram}
\end{figure*}

The quantity 
\be\label{Nc} A_\alpha(\lambda;x)=\prod_{i=1}^{\ell(\lambda)}\prod_{j=1}^{\lambda_i}[x-c_\alpha(i,j)]\ee
is the $\alpha$-content polynomial of $\lambda$. When $x=N$ is an integer, this coincides with the value of the Jack polynomial at the identity,
\be A_\alpha(\lambda;N)=J_\lambda^{(\alpha)}(1^N)=[N]_\lambda^{(\alpha)}.\ee
The smallest power of $N$ in $A_\alpha(\lambda;N)$ is precisely $D_\alpha(\lambda)$, by definition. Its coefficient is the product of all $\alpha$-contents that are not zero. Therefore, we have
\be [N]_\theta^{(\alpha)}= t_\alpha(\theta) N^{D_\alpha(\theta)}+O(N^{D_\alpha(\theta)+1}).\ee

When the diagram of $\lambda$ is contained in the diagram of $\theta$, the skew shape $\theta/\lambda$ exists and is the complement. An example is shown Figure \ref{diagram} in which $\theta=(4,4,2,2,1)$ and $\lambda=(2,2,1)$. The function $t_\alpha(\theta/\lambda)$ is then the product of all non-zero $\alpha$-contents in $\theta/\lambda$.

\end{document}